\documentclass[floats,floatfix,
  amssymb,aps, prd,superscriptaddress,
  nofootinbib,preprintnumbers,notitlepage,
  twocolumn, 10pt]{revtex4-1}
\usepackage{amssymb, amsmath, verbatim, mathtools, needspace, enumitem, etoolbox, graphicx, physics, microtype, afterpage, bm}
\usepackage[dvipsnames]{xcolor}
\usepackage[T1]{fontenc}
\usepackage[utf8]{inputenc}
\usepackage{tabularx,multirow,booktabs}
\usepackage{float}
\usepackage{pifont}
\usepackage{lmodern}
\usepackage[normalem]{ulem}

\allowdisplaybreaks
\usepackage{tikz}
\usepackage{framed}
\usepackage{hyperref}
\definecolor{bluscuro}{rgb}{0.15, 0.2, .85}
\hypersetup{colorlinks,
  linktocpage,
  citecolor=bluscuro, 
  linkcolor=Red, 
  urlcolor=Mulberry}
\usepackage[capitalize]{cleveref}

\makeatletter
\newcommand{\subsetsim}{\mathrel{\mathpalette\subset@sim\relax}}
\newcommand{\subset@sim}[2]{%
  \vtop{\offinterlineskip\m@th
    \ialign{\hfil##\cr
      $#1\subset$\cr\noalign{\kern0.5pt}\scalebox{0.9}{$#1\sim$}\cr
    }%
  }%
}
\makeatother

\newcommand{\GeV}{\,\text{GeV}}

\newcommand{\keV}{\,\text{keV}}

\newcommand{\g}{\,\text{g}}
\newcommand{\gs}{g_\star}

\newcommand{\be}{\begin{equation}}
\newcommand{\ee}{\end{equation}}

\def\PBH{{\rm  PBH}}

\newcommand{\MP}{M_{\rm P}}
\newcommand{\parfrac}[2]{\left(\frac{#1}{#2}\right)}

\newcommand{\cgw}{\text{\scriptsize{GW,cos}}}
\newcommand{\MBH}{M_\text{PBH}}
\newcommand{\TBH}{T_\text{PBH}}
\newcommand{\nSM}{n_\text{\scriptsize SM}}
\newcommand{\mDM}{m_\text{DS}}
\newcommand{\epsRH}{\epsilon_\text{\scriptsize RH}}
\newcommand{\fNL}{f_\text{\scriptsize NL}}
\newcommand{\dpbh}{\delta_\text{\scriptsize PBH}}
\newcommand{\td}{{\rm d}}
\renewcommand{\dd}{\td}
\newcommand{\HI}{H_\text{\scriptsize I}}
\newcommand{\He}{H_\text{e}}
\newcommand{\Neff}{N_\text{eff}}
\newcommand{\Tmax}{T_\text{max}}
\newcommand{\TRH}{T_\text{\scriptsize{inst.\,RH}}}
\newcommand{\ODM}{\Omega_\text{\scriptsize{DM}}}

\definecolor{rossos}{cmyk}{0,1,1,0.55}
\definecolor{bluscuro}{rgb}{0.15, 0.2, .85}
\definecolor{bluchiaro}{cmyk}{1,.3,0.,0.1}
\definecolor{TLGreen}{RGB}{50, 164, 49}
\definecolor{TLOrange}{RGB}{231,180,22}
\definecolor{TLRed}{RGB}{204,50,50}

\newcommand{\unipd}{Dipartimento di Fisica e Astronomia ``G. Galilei'', Università degli Studi di Padova, via Marzolo 8, I-35131 Padova, Italy}
\newcommand{\infnpd}{INFN, Sezione di Padova, via Marzolo 8, I-35131 Padova, Italy}

\begin{document}

\title{Isocurvature Constraints on Dark Matter from Evaporated Primordial Black Holes}
\author{Gabriele Franciolini}
\email{gabriele.franciolini@unipd.it}
\affiliation{\unipd}
\affiliation{\infnpd}
\author{Davide Racco}
\email{davide.racco@uniroma3.it}
\affiliation{Dipartimento di Matematica e Fisica, Università degli Studi Roma Tre and INFN Sezione di Roma Tre, Via della Vasca Navale 84, 00146, Rome, Italy}
\affiliation{Physik-Institut, Universit\"at Z\"urich, Winterthurerstrasse 190, 8057 Z\"urich, Switzerland}
\affiliation{Institut f\"ur Theoretische Physik, ETH Z\"urich,Wolfgang-Pauli-Str.\ 27, 8093 Z\"urich, Switzerland}

\date{\today}


\begin{abstract}
We revisit the scenario in which stable particles of a dark sector are produced through the complete evaporation of light primordial black holes (PBHs) formed in the early Universe. We investigate in detail the role of isocurvature perturbations that may arise in this framework. PBHs inherit Poisson fluctuations on unobservable small scales at formation; however, in the presence of primordial non-Gaussianity that couples long- and short-wavelength modes, these fluctuations can source isocurvature perturbations on cosmological scales. Such perturbations are unavoidably transferred to the dark sector particles emitted via Hawking evaporation. We highlight the potential impact of isocurvature constraints on dark sector particles produced through PBH evaporation.
Along the way, we re-assess the constraints on this scenario arising from the overproduction of dark matter (DM), accounting for both PBH evaporation and gravitational production (freeze-in) during (after) inflation, as well as bounds from warm DM and the overproduction of scalar-induced gravitational waves.
\end{abstract}

\maketitle
\preprint{CERN-TH-2026-018} 
\preprint{ZU-TH 09/26}
\tableofcontents

\section{Introduction}

Primordial black holes (PBHs) \cite{Byrnes:2025tji} are an interesting class of objects potentially formed in the early Universe from the collapse of overcritical perturbations. 
PBH may form in a variety of scenarios, from enhanced small-scale curvature perturbations, non-standard thermal histories and possibly phase transitions, and reheating, see e.g.~\cite{Ozsoy:2023ryl,Flores2025} for recent reviews.
While most studies focus on the mass range in which PBHs are stable over cosmological timescales and their Hawking evaporation is negligible~\cite{Hawking:1974rv,Hawking:1974sw}, many scenarios consider PBHs formed in the early Universe as a source of particle production during their complete evaporation prior to BBN. In this process, all particle species are produced due to the universal nature of gravitational interactions. In particular, dark sector (DS) particles would also be generated, potentially accounting for a significant fraction of the dark matter (DM) observed today. In this work, we take consistently into account various constraints on this scenario, and in particular we investigate the role of isocurvature perturbations.

We aim to revisit and extend previous studies on the constraints associated with PBHs producing particle DM. The coexistence of these two new physics phenomena leads to rich phenomenology, and several effects could render this scenario incompatible with standard cosmological observations. In particular, DM particles produced by Hawking evaporation of PBHs could potentially overclose the Universe. In some regions of parameter space, this may exceed the inevitable DM production from inflationary gravitational mechanisms, also leading to overclosure. Furthermore, the abundance of ultralight PBHs may be constrained by gravitons produced either through PBH evaporation or as scalar-induced gravitational waves (GWs), which can surpass current upper limits on stochastic GW backgrounds. Additionally, DM particles generated via evaporation might avoid overclosure but remain relativistic for too long, violating warm DM bounds. While these effects have been studied extensively (but not all at once) in the literature \cite{Lennon:2017tqq,Morrison:2018xla,Hooper:2019gtx,Gondolo:2020uqv,Masina:2020xhk,Bernal:2020bjf,Bernal:2020ili,Masina:2021zpu,Cheek:2021odj, Cheek:2021cfe, Cheek:2022dbx, Cheek:2022mmy,Paul:2026ovy}, we aim to further explore the possibility that DM produced by evaporation could carry isocurvature perturbations.

PBHs possess a characteristic Poisson noise in their distribution, which provides an unavoidable source of isocurvature \cite{Ali-Haimoud:2018dau,Desjacques:2018wuu,Inman:2019wvr,DeLuca:2020jug}. However, this is typically predicted at the scale of the mean PBH separation at high redshift, making it inaccessible to CMB observations. In the presence of primordial non-Gaussianity (NG) that couples long and short modes, however, isocurvature perturbations may be imprinted on the dark sector at cosmological scales. In this work, we focus in particular on the implications of these NG-induced isocurvature perturbations.\footnote{In this work we assume that Hawking evaporation is not modified by the suggested \textit{memory burden} effect \cite{Dvali:2018xpy, Dvali:2020wft, Balaji:2024hpu, Dvali:2024hsb, Chianese:2024rsn,  Chaudhuri:2025asm, Dondarini:2025ktz, Dvali:2025ktz, Chianese:2025wrk, Montefalcone:2025akm}.}

We review in \cref{sec:DM from PBH} the production of stable particles of the DS (which we occasionally refer to as DM, although they might not constitute the totality of DM) from the gravitational channels of Hawking evaporation of PBHs, gravitational production during inflation or gravitational freeze-in after inflation.
\cref{sec:WDM and SIGW} discusses other constraints on this scenario due to the warmness of the evaporated DM, or the overproduction of gravitons from Hawking evaporation or scalar-induced tensor perturbations.
We focus in \cref{sec:Iso PBH NG} on the large-scale isocurvature perturbations which PBHs can possess in presence of squeezed NG, to analyse for the first time how this isocurvature is inherited by the evaporated DS particles, which are then constrained to a sub-population of DM.
\cref{sec:results} contains the summary plots of our analysis, and we conclude in \cref{sec:conclusions}.
Finally, in \cref{app:PBH_formation} we review the computation of the PBH abundance.

\section{Dark matter from PBH evaporation and gravitational production}
\label{sec:DM from PBH}

In this section we describe the basic formalism to describe Hawking evaporation of PBHs in a cosmological context.
We adopt the public code \cite{code_frisbee} (based on \cite{Cheek:2021odj, Cheek:2021cfe, Cheek:2022dbx, Cheek:2022mmy}) to compute the relic abundance of DM particles produced by PBHs.
When tracking the evolution of the PBH population, we neglect the possible role of accretion and the presence of a thermal bath, see e.g. \cite{Kalita:2025fcs,Haque:2026vvp,Chaudhuri:2026zyx}. 

\subsection{Basics of the PBH evaporation}\label{sec:PBH_evap}

In the following, we focus on non-rotating (Schwarzschild) black holes, as expected in most PBH formation scenarios \cite{Mirbabayi:2019uph,DeLuca:2019buf}. 
The extension to rotating (Kerr) black holes is left for future work. 
As shown by Hawking~\cite{Hawking:1974rv,Hawking:1974sw}, a Schwarzschild BH of mass $\MBH$ emits particles with a thermal spectrum at temperature
\begin{equation}
\TBH = \frac{1}{8\pi G \MBH} \sim 1.06~{\rm GeV} \left(\frac{10^{13}~{\rm g}}{\MBH}\right),    
\end{equation}
where $G$ is the gravitational Newton constant. 

The emission rate of a species $i$ (with mass $\mu_i$, spin $s_i$, and $g_i$ degrees of freedom) per unit time and momentum interval $[p, p + \dd{p}]$ can be written as
\be\label{eq:BHrate}
\frac{\dd^2 \mathcal{N}_{i}}{\dd p\,\dd t} = \frac{g_i}{2\pi^2} \frac{\sigma_{s_i}(\MBH, \mu_i, p)}{\exp\left[E_i(p)/\TBH\right] - (-1)^{2s_i}} \frac{p^3}{E_i(p)},
\ee
with $E_i(p) = \sqrt{\mu_i^2 + p^2}$ and $\sigma_{s_i}$ the absorption cross-section. This rate governs both the BH mass loss and the distribution of emitted particles. The cross-section $\sigma_{s_i}$ (or equivalently the greybody factor $\Gamma_{s_i} \equiv \sigma_{s_i} p^2 / \pi$) accounts for potential back-scattering due to gravitational or centrifugal barriers~\cite{Hawking:1974rv,Hawking:1974sw,Page:1976df,Page:1977um}. While often neglected, recent studies~\cite{Auffinger:2020afu,Masina:2021zpu} offer a detailed inclusion of these effects. Here we follow the recent analysis of Ref.~ \cite{Cheek:2021odj}.

As a consequence of Hawking emission, BHs lose mass over time.
The total mass loss rate is obtained by summing \cref{eq:BHrate} over all particle species and integrating over momentum~\cite{MacGibbon:1990zk,MacGibbon:1991tj}:
\begin{equation}
\label{eq:dMBHdt}
\frac{\dd \MBH}{\dd t} 
= -\sum_i \int_0^\infty 
\hspace{-.7em}
E_i \frac{\dd^2 \mathcal{N}_i}{\dd p\, \dd t} \dd p = -\varepsilon(\MBH) \frac{\MP^4}{\MBH^2},%
\end{equation}
where $\MP = G^{-1/2}$ is the unreduced Planck mass. The function $\varepsilon(\MBH)$ encodes the evaporation efficiency at a given mass and is defined as
\begin{equation}
\varepsilon(\MBH) \equiv \sum_i g_i\, \varepsilon_i(z_i),
\end{equation}
with $z_i = \mu_i / \TBH$ and
\begin{equation}
\varepsilon_i(z_i) = \frac{27}{8192\pi^5} \int_{z_i}^\infty \frac{\psi_{s_i}(x)\, (x^2 - z_i^2)}{\exp(x) - (-1)^{2s_i}}\, x\, \dd x,
\end{equation}
where $x = E_i / \TBH$ and the dimensionless greybody factor is defined as
$
\psi_{s_i}(E) \equiv \sigma_{s_i}(E)/({27\pi G^2 \MBH^2}).
$
The explicit forms of $\varepsilon_i(z_i)$ can be found e.g.~in \cite[App.~A]{Cheek:2021odj}. 

We can now compute the momentum-integrated emission rate $\Gamma_{{\rm BH} \to i}$ by integrating \cref{eq:BHrate} over momentum $p$, to find
\begin{equation}\label{eq:Hwtotrate}
\Gamma_{\text{\scriptsize{PBH}} \to i} 
= 9.8\cdot 10^{29}
{\rm s^{-1}}
g_i  \left(
\frac{10^5{\rm g}}{\MBH}
\right) \left(
\frac{\Psi_i(z_i)}{0.897}
\right),
\end{equation}
where
\begin{equation}
\Psi_i(z_i) \equiv \int_{z_i}^{\infty} 
\frac{\psi_{s_i}(x)\left(x^2 - z_i^2\right)}{\exp(x) - (-1)^{2 s_i}} \, \mathrm{d}x \, .
\end{equation}
Overall, we can estimate the time by which the BH has evaporated
\begin{equation}\label{eq:tev}
t_{\rm ev} \sim 4 \times 10^{-28} ~{\rm s} 
\left( \frac{\MBH^{\rm in}}{\rm g} \right)^3.
\end{equation}
As a consequence, we see that PBHs with initial mass $\MBH^{\rm in} \gtrsim 10^{15}~{\rm g}$ have lifetimes longer than the current age of the Universe.

\subsection{%
PBH impact on cosmological history
}

For simplicity, we assume that PBHs form during radiation domination with an initial mass proportional to the horizon mass~\cite{Carr:2020xqk}
\begin{align}\label{eq:Min}
\MBH^{\rm in} 
= \frac{\gamma}{2}
\frac{\MP^2}{H_\times} 
&= 
8.0~{\rm g }
\left ( \frac{\gamma}{0.6} \right ) 
\left ( \frac{H_\times}{10^{13}{\rm GeV}} \right )^{-1}
\nonumber \\& 
\sim 
24~{\rm g }\left( \frac{t_i}{10^{-37} {\rm s} }\right),
\end{align}
where $H_\times$ ($\sim 1/(2 t_i)$) is the Hubble rate at Hubble crossing of the dominant mode for PBH production, and $\gamma \sim 0.6$ characterizes the collapse efficiency~\cite{Franciolini:2023osw}.\footnote{Notice in the actual computation we take into account all the details of the critical collapse, as discussed in \cref{app:PBH_formation}.}
We also assume for simplicity of notation that the PBH mass function is very narrow, and approximate it as monochromatic, but one can easily extend what follows to an extended mass distribution.
The PBH energy density at formation is encoded in
\begin{equation}\label{eq:betap}
\beta \equiv \gamma^{1/2} \left( \frac{g_\star(T_{\rm in})}{106.75} \right)^{-1/4} \frac{\rho_{\rm PBH}^{\rm in}}{\rho^{\rm in}},
\end{equation}
with $T_{\rm in}$ the plasma temperature at formation.

In general, the different scaling of energy density in PBHs and radiation with the scale factor $a$ can enhance the PBH abundance $\beta$ by up to 25 orders of magnitude, depending on the initial PBH mass. 
Consequently, even a tiny initial fraction may grow sufficiently to trigger an early PBH-dominated (ePBHD) epoch, depending on whether the initial fraction is larger then $\beta_c$, defined below. 
We denote the ePBHD start time as $t_d$, with an associated scale factor $a_d = a_i / \beta_i$. Since their energy density fraction grows linearly with $a$, the time when PBHs first dominate is given by
$t_d = t_i/\beta_i^2$.
Given that the PBH survives for $t_{\rm ev}$ given by \cref{eq:tev}, the  condition for this to happen is
\begin{equation}
\beta_i > \beta_{\rm c} = \left( \frac{t_i}{t_{\rm ev}} \right)^{1/2} \simeq 1.6 \times 10^{-6} \left( \frac{M^{\rm in}_{\rm PBH}}{\rm g} \right)^{-1}.
\label{eq:beta_crit}
\end{equation}
If this condition is met, one goes through an epoch of ePBHD, and the subsequent energy content after PBH evaporation is solely dictated by the grey-body factors controlling the Hawking emision. 

Alongside the evolution of PBHs, the evolution of the Hubble rate is controlled by the Friedmann equation, in which we account for the possible various components
corresponding to standard model (SM), dark sector (DS) and PBHs. 
PBHs affect both the cosmic expansion and the energy budget via Hawking evaporation, which -- owing to the universality of gravitational interactions -- involves all DS particles.
In turn, the cosmological evolution of the DM abundance produced from PBH evaporation depends on two important effects: {\it i)} At which stage in the evaporation  the BH temperature  $\TBH$ rise above $\mDM$, and {\it ii)} whether PBHs evaporate during radiation or matter domination~\cite{Lennon:2017tqq,Morrison:2018xla,Hooper:2019gtx,Gondolo:2020uqv,Masina:2020xhk,Bernal:2020bjf,Bernal:2020ili,Masina:2021zpu}, which affects the subsequent dilution rate.  All these effects are accounted for in our computation, based on the public code \cite{code_frisbee}.

The resulting DM abundance today is given by entropy conservation
\begin{equation}
\Omega_{\rm DM} = 
\frac{1}{\rho_{\rm crit}^0} 
\left (\frac{g_{\star s}(T_0)}{g_{\star s}(T_{\rm ev})} \right ) 
\left (\frac{T_0^3}{ T_{\rm ev}^3} \right )
\rho_{\rm DM}^{\rm ev},
\end{equation}
with $T_0$ the present CMB temperature and $T_{\rm ev}$ being the temperature of complete evaporation of the PBHs. 
The unavoidable Hawking emission process can lead to the overproduction of the total DM (or of the allowed fraction of warm DM, see the following section). The bound is driven by the heaviest stable particle in the DS, whose mass will be denoted by $\mDM$. 

Now that we have established the basic ingredients of PBH evolution, together with the associated cosmological history described in this section, we turn our attention to complementary DM production channels, which are inevitably present irrespectively of the existence of PBHs.

\subsection{Gravitational production during and after inflation}
\label{sec:DM from GPP}

An unavoidable production channel for DM is gravitational production. 
We can distinguish the following contributions.

\begin{figure*}[t]\centering
\includegraphics[width=\textwidth]{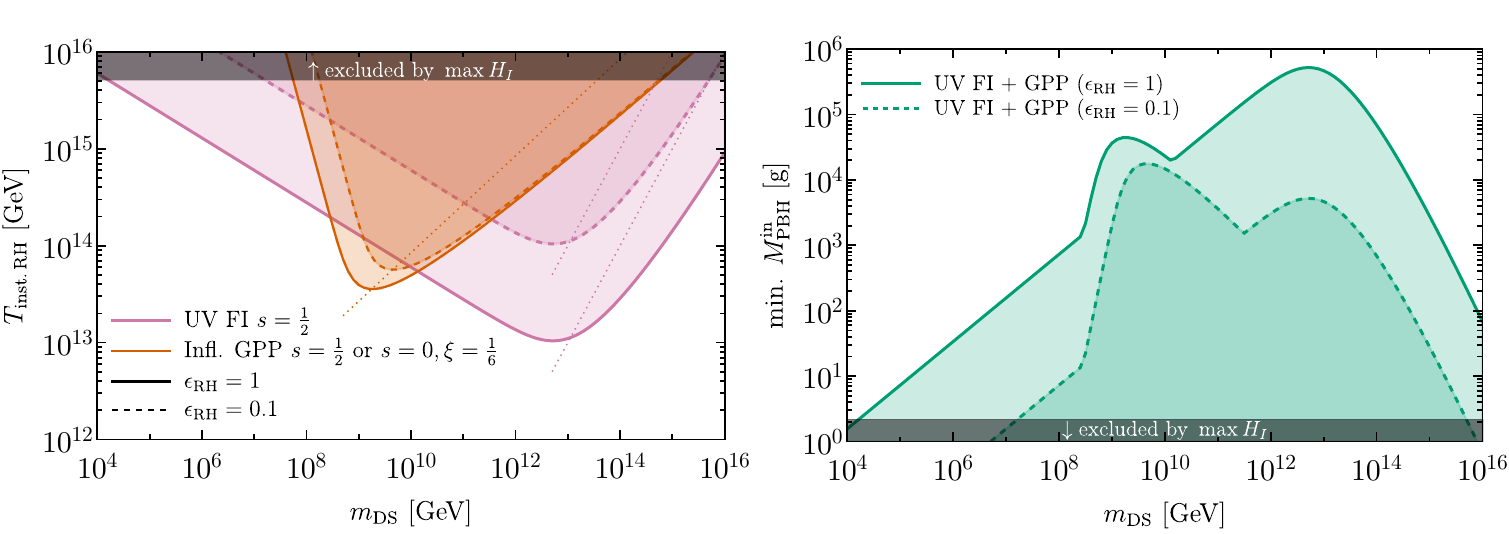}
\caption{
\textit{Left}: Parameter space where a DM particle is overproduced by inflationary production (considering a fermion or minimally-coupled scalar, in orange) or by UV freeze-in (in pink), for two values of $\epsRH= \Tmax/\TRH$. Below the dotted lines, these yields are Boltzmann-suppressed.
\textit{Right}: translation of these bounds to the minimum PBH mass that is achievable for a given $\mDM$.}
\label{fig:DM GPP and UV FI}
\end{figure*}

\paragraph{Graviton-mediated freeze-in}
The scattering of SM particles in the thermal bath can produce, via a graviton mediator, a pair of DM particles, without assuming any other interaction.
As the gravitational coupling grows with energy (or equivalently, the effective operator for this process is irrelevant in the IR), this process is dominated by the UV, i.e.~the largest temperature $\Tmax$ after reheating.
The highest possible $\Tmax$ is achieved for maximal efficiency of reheating, corresponding to the temperature $\TRH$ of instantaneous reheating.
We denote 
\begin{equation}
    \epsRH \equiv \frac{\Tmax}{\TRH} \,.
\end{equation}
We adopt the results of Ref.~\cite{Bernal:2018qlk} for the rate of DM production, taking for concreteness the case of fermion DM (with an $\mathcal O(0.1)$ factor of difference for scalars, see Table I of \cite{Bernal:2018qlk}). 
The production rate is
\begin{equation}
    \Gamma =\nSM^2\sigma \,, \quad \sigma \simeq 1.1\cdot 10^{-6}\frac{(8\pi)^2\Tmax^2}{\MP^4}\,,
\end{equation}
with $\nSM = \tfrac{\zeta(3)}{\pi^2}\gs\Tmax^3$, 
and we approximate the integrated energy density to the one produced in a Hubble time after the end of inflation, $\rho_{\rm DM}(\Tmax) \simeq \He^{-1}\Gamma(\Tmax)$.

\paragraph{Gravitational production during inflation}
For a scalar, generically one expects overclosure from ``misalignment'' contribution, that is the zero-mode of the scalar field, because the potential energy of a scalar field does not redshift for super-Hubble modes.
For this reason, a fundamental scalar with mass lighter than the Hubble rate during inflation, if it does not thermalise with the bath at some time after inflation, generically overcloses the universe (see e.g.~\cite{Kolb:2023ydq}).
This changes if the scalar is conformally coupled ($\xi = \frac 16$), as in that case the equation of motion for the field is conformally invariant up to the mass term. In this special case, the gravitational production of the massive scalar resembles the one of a massive fermion \cite{Chung:1998zb,Chung:1998ua}.
This point should be kept in mind whenever considering gravitational (or Hawking) production of a secluded species, and for this reason in the left panel of  \cref{fig:DM GPP and UV FI}, the orange line can be assumed equivalently to refer to a single helicity state of a massive fermion or scalar that is conformally coupled.
The final abundance of such species is (see e.g.~\cite{Kolb:2023ydq})
\begin{equation}
\ODM h^2 \simeq 0.12 \cdot \parfrac{\mDM}{10^{11} \GeV}^2 \parfrac{\Tmax}{10^{9} \GeV} e^{-2\frac{\mDM}{\He}},
\end{equation}
where in the Boltzmann suppression factor we approximate $\He$ as roughly constant during inflation, and we assume no significant contribution to the energy density from the reheating stage \cite{Racco:2024aac}.

\section{Indirect bounds}
\label{sec:WDM and SIGW}

\subsection{Warm dark matter}
\label{sec:WDM}
An important constraint on DM produced purely via Hawking evaporation arises from the warm dark matter (WDM) bound. 
Owing to the non-thermal phase space distribution of Hawking-evaporated particles, and their injection at cosmologically late times, they damp the high-$k$ tail of the matter power spectrum  due to free-streaming.
The matter perturbations measured with Lyman-$\alpha$ observations \cite{Irsic:2017ixq} can be recast as a bound $m_\text{WDM}>3\keV$ on the mass of a thermal WDM candidate, which allows for a quick estimate of the corresponding bound on Hawking-evaporated particles via the calculation of their average momentum (as e.g.~in \cite{Cheek:2021odj}).\\
As a more refined treatment, the suppression in the matter power spectrum can be parameterised as in \cite{Viel:2005qj} with a power-law, which can then be fit against numerical solution of the linear Boltzmann equations (e.g.~with \textsc{class} \cite{CLASSI,CLASSII,CLASSIV}) with the DM phase-space distribution given as an input \cite{Baldes:2020nuv,Auffinger:2020afu}. 
The final result is that the WDM constraints are stronger for lighter DM (which remains relativistic for a longer time), and they exclude 
\begin{equation}
\beta \gtrsim 0.016 \,\beta_c 
\end{equation}
for a fermionic Hawking-evaporated DM \cite{Baldes:2020nuv, Masina:2020xhk}. The bound is almost independent from the particle spin, except for the spin-1 case with $\beta < 0.03\,\beta_c$ \cite{Masina:2020xhk}.
We mark the corresponding excluded region with a thick shading in \cref{fig:DM_PBH}, only along the lines where the Hawking-evaporated particles are the totality of DM, following the choice of \cite{Cheek:2021odj}. 
Notice indeed that the comparison with the parameterisation of the matter PS of \cite{Viel:2005qj} requires that  the Hawking-evaporated particles are the only DM component. In other words, the bound inevitably depends on three variables ($\MBH^{\rm in}, \beta, \mDM$) and cannot be easily compressed in the commonly-chosen two-dimensional parameter space $(\MBH^{\rm in},\beta)$.\\
It would be very interesting to extend these studies in two directions (for some progress in this direction, see \cite{Chen:2025gnm}). 
First, it would be important to include another DM component on top of the Hawking-evaporated particles, in order to cover the region of parameter space below the 100\%-DM lines of \cref{fig:DM_PBH} (for a given $\mDM$). 
Secondly, it could be necessary in that case to enforce a full $\chi^2$ test against the data, as one cannot rely on the parameterisation of \cite{Viel:2005qj}.
We leave these interesting extensions to future work.

\subsection{Bounds from gravitational waves}\label{sec:BBNbound}

BBN and CMB observations tightly constrain the total energy density stored in relativistic species beyond the SM. This is conventionally parametrized in terms of the effective number of relativistic degrees of freedom, $\Delta N_{\rm eff} \equiv N_{\rm eff} - N_{\rm eff}^{\rm SM}$, where $N_{\rm eff}^{\rm SM} \simeq 3.043$ \cite{Cielo:2023bqp} accounts for the contribution of the three active neutrino species. Any additional relativistic component present during BBN or recombination, including a SGWB, contributes positively to $\Delta N_{\rm eff}$ and is therefore constrained by observations. 
Measurements of the CMB~\cite{Planck:2018vyg} and Baryon Acoustic Oscillations (BAO) constrain $\Delta \Neff\leq 0.28$ at $95\%$ C.L.
These limits can be translated into an upper bound on the total GW energy density at BBN. 
Specifically, the total (integrated) GW abundance is $\Omega_\cgw h^2\simeq 1.6\cdot 10^{-6} \left(\Delta N^{\rm GW }_{\text{eff}}/0.28\right)$~ \cite{Caprini:2018mtu}. 

In the scenarios under consideration, involving ultra-light PBHs, the existence of a GW component is unavoidable, as PBHs necessarily emit relativistic degrees of freedom in the form of gravitons even in the absence of any BSM physics. Below we discuss the two irreducible contributions to $\Delta N_{\rm eff}$ associated with PBHs.

\subsubsection{Gravitons from Hawking evaporation}
\label{subsec:HawkingNeff}

PBHs emit all particle species lighter than their Hawking temperature, including gravitons. The fraction of the total PBH mass converted into gravitons is determined by the Hawking evaporation process and, quantitatively, by the corresponding gray-body factors, see Sec.~\ref{sec:PBH_evap}. If PBHs come to dominate the energy density of the Universe prior to evaporation, the subsequent reheating is entirely controlled by Hawking emission. In this regime, the ratio between the graviton energy density and the SM thermal bath after reheating becomes independent of the initial PBH abundance $\beta$ and is fixed solely by the gray-body factors. Assuming an epoch of spinless PBH domination, the contribution to $\Delta N_{\rm eff}$ from gravitons was computed to be $\Delta N_{\rm eff} = 0.003 \divisionsymbol 0.006$ \cite{Hooper:2019gtx}, where the uncertainty depends on the PBH mass, and corresponding initial PBH temperature affecting which SM degrees of freedom can be radiated. We conservatively neglect the possible existence of additional BSM degrees of freedom, which would further deplete the graviton final abundance, as well as possible changes due to a non-trivial BH spin evolution (see e.g. \cite{Calza:2021czr}).

As a consequence, current bounds on $\Delta N_{\rm eff}$ do not provide constraining power on $\beta$ once $\beta$ exceeds the critical value $\beta_c$ required for PBH domination. The resulting $\Delta N_{\rm eff}$ from direct graviton emission is too small to be constrained by present BBN and CMB measurements. Future experiments such as CMB-S4 may significantly improve the sensitivity to $\Delta N_{\rm eff}$ and potentially probe this irreducible contribution provided graviton emission is boosted assuming spinning PBHs \cite{Hooper:2020evu}. 

\subsubsection{Induced gravitational waves}
\label{subsec:InducedGWs}

A second, independent and unavoidable contribution to the GW background arises from second-order induced GWs sourced by PBH-induced small-scale isocurvature perturbations. During the PBH-dominated era, these perturbations grow and, after complete PBH evaporation, are converted into adiabatic curvature perturbations, which in turn source GWs at second order. Following Ref.~\cite{Domenech:2020ssp}, the resulting GW energy density contributes to $\Delta N_{\rm eff}$ and is subject to the BBN bound.

Imposing $\Omega_{\rm GW,BBN} \lesssim 0.05$ leads to an upper limit on the initial PBH abundance,
\begin{equation}
\beta \lesssim 1.1 \times 10^{-6}
\left( \frac{M^{\rm in}_{\rm PBH}}{10^4\,{\rm g}} \right)^{-17/24},
\label{eq:betaBBNbound}
\end{equation}
in agreement with Refs.~\cite{Inomata:2020lmk,Domenech:2020ssp}. The precise amplitude of the induced GW signal depends on the rapidity of the transition between PBH domination and radiation domination, and therefore indirectly on the PBH mass distribution. In this work we adopt the treatment of Ref.~\cite{Domenech:2020ssp} and leave a more refined analysis of these effects for future investigation. We also neglect possible additional GW emission from PBH mergers, which could further enhance the signal (see e.g.~\cite{Hooper:2020evu}).

\section{Isocurvature perturbations imprinted by PBHs}
\label{sec:Iso PBH NG}

PBHs represent a population of discrete objects and therefore act as point-like tracers of the overall dark matter distribution. The associated PBH overdensity can be written as
\begin{equation}
\delta_{\rm PBH}(\vec{x})=\frac{1}{n_{\rm PBH}}\sum_i \delta_D(\vec{x}-\vec{x}_i)-1 \, ,
\end{equation}
where $\delta_D(\vec{x})$ denotes the three-dimensional Dirac delta function and $n_{\rm PBH}$ is the mean comoving PBH number density. 
The index $i$ labels the initial spatial positions of PBHs.
In terms of the PBH initial mass fraction, this quantity can be expressed as
\begin{equation}
n_{\rm PBH}
\simeq \left (\frac{1 }{100 \, {\rm km}}  \right)^3
\left ( \frac{\beta}{10^{-10} }\right )
\left ( \frac{M^{\rm in}_{\rm PBH}}{10^{4} {\rm g} }\right )^{-3/2},
\end{equation}
Since PBHs constitute a discrete set of objects, their spatial distribution is subject to Poisson fluctuations. The typical fluctuation within a comoving volume $V$ scales as $\delta_{\rm PBH}\sim N^{-1/2}$, where $N=n_{\rm PBH}V$ is the expected number of PBHs inside the volume. This immediately defines a characteristic comoving length scale associated with discreteness effects,
\begin{equation}
\lambda_{\rm P}\sim n_{\rm PBH}^{-1/3} \, ,
\end{equation}
which corresponds to the average separation between neighboring PBHs. This length determines the typical scale at which PBH Poisson noise sources isocurvature perturbations in the DM distribution. Perturbation modes with wavelengths $\lesssim \lambda_{\rm P}$ are therefore dominated by discreteness fluctuations, whereas longer-wavelength modes experience a suppressed contribution. Given the very small masses considered in this work, the isocurvature modes associated to Poisson fluctuations are unobservable at large scales. 

Furthermore, if PBHs represent only a subdominant fraction of DM, gravitational clustering among PBHs evolves too slowly to significantly alter the initial Poisson-induced fluctuations over timescales comparable to the PBH evaporation time. Consequently, the resulting estimate of the DM isocurvature perturbations remains largely insensitive to the later nonlinear PBH clustering, making this argument robust.%
\footnote{In deriving the isocurvature perturbations, we neglect possible diffusion effects. Since the DM particles can be relativistic at high redshift, free streaming may partially erase isocurvature fluctuations. However, observational constraints require particle DM to be sufficiently cold at low redshift, implying that this suppression must be small on CMB scales. We therefore neglect this effect in our analysis and leave a dedicated quantitative assessment to future work.}

In the following, we show that initial isocurvature perturbations can be enhanced beyond Poisson noise, but only if specific non-Gaussian perturbations are assumed, correlating short (PBH scale) and long (CMB scale) modes.

\subsection{Gaussian perturbations}
We adopt the peak-background split picture, where perturbations are separated into short ($\zeta_s$) and long ($\zeta_l$) modes~\cite{Sheth:1999mn}. Using the threshold on the compaction function $\mathcal{C} > {\cal C}_{\rm th}$ to model PBH formation, see App.~\ref{app:PBH_formation} for more details, we define the non-Poisson fluctuation as (e.g. \cite{Tada:2015noa})
\begin{align}
\dpbh({\vec x}) & = \frac{P\left(\mathcal{C} > \mathcal{C}_{\rm th}|\zeta_l(\vec x)\right)}{P\left(\mathcal{C} > \mathcal{C}_{\rm th}\right)}-1
\nonumber \\
& 
\simeq\left. 
\frac{\partial {\cal C}}{\partial \zeta_l}
\frac{
\partial\ln P\left(\mathcal{C} > \mathcal{C}_{\rm th}|\zeta_l(\vec x)\right)
}{
\partial {\cal C}
}\right|_{\zeta_l=0}\zeta_l (\vec x)
\nonumber \\
& 
\simeq 
\frac{\partial {\cal C}}{\partial \zeta_l}
\frac{\nu}{\sigma_{\rm c}} \zeta_l (\vec x).
\end{align}
In the last step, we assumed the high-peak limit $\nu \equiv {\cal C}_{\rm th}/{\sigma_{\rm c} } \gg 1$, which gives the usual scale-independent bias term \cite{Tada:2015noa}. However, although $\nu$ can be large, the prefactor is strongly suppressed. 
Neglecting non-linearities at this stage, as they are irrelevant for the current argument, we approximate the compaction function \cref{eq:CompactionFull} as ${\cal C} \sim {\cal C}_1 \sim -4 (k_l r_m) \zeta/3$, when expressed in momentum space. 
As the typical size of a perturbation collapsing into a PBH is 
of the order of the Hubble radius at $k_s$ crossing, 
$r_m \sim {\cal O}(3)/k_s$ \cite{Musco:2020jjb}, one finds a suppression of order $k_l / k_s \ll 1$.%
\footnote{The computation of the PBH abundance via a threshold on the curvature perturbation (rather than its Laplacian $\delta$, or more accurately on $\cal C$) leads to erroneous conclusions, as an unphysical constant $\zeta$ mode may alter the collapse probability \cite{Young:2014ana}, and the consequent isocurvature perturbations \cite{Kim:2025kgu}.} 
As we will see in the following, including local non-Gaussianity can avoid this suppression, by introducting a strong scale-dependent bias.

\subsection{Impact of non-Gaussianities}

The strong conclusion about the lack of isocurvature on long scales can be avoided if primordial non-Gaussianity (NG) in $\zeta$ significantly affects PBH formation~\cite{Young:2013oia, Bugaev:2013vba, Young:2014ana, Nakama:2016gzw, Byrnes:2012yx, Franciolini:2018vbk, Yoo:2018kvb,Kawasaki:2019mbl, Riccardi:2021rlf,Taoso:2021uvl, Biagetti:2021eep, Kitajima:2021fpq, Hooshangi:2021ubn, Meng:2022ixx, Young:2022phe, Escriva:2022pnz, Hooshangi:2023kss, Ianniccari:2024bkh, Ferrante:2022mui}, correlating the PBH distribution on very large scales. In this work, we will be agnostic about the specific model that can realize this kind of NG correlation, and adopt a general parametrization where 
$\zeta = F(\zeta_{\rm g}),$
in terms of the Gaussian component of the curvature perturbation $\zeta_{\rm g}$.

\begin{figure}[t]
    \centering
    \includegraphics[width=0.49\textwidth]{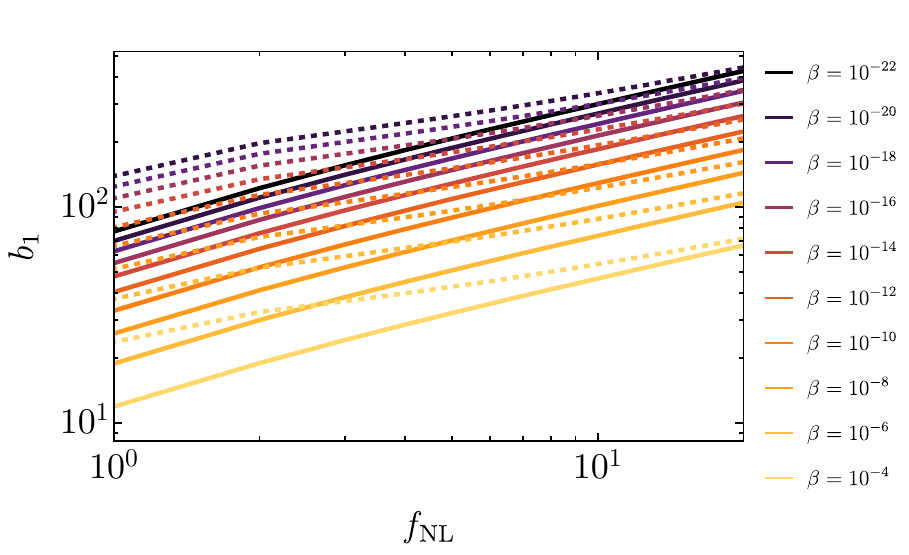}
    \caption{
    Variation of the linear bias $b_1$ as a function of the local-type non-Gaussianity parameter $\fNL$ for several values of $\log_{10}\beta$ (marked with different colors as shown in the legend). Solid (dashed) lines corresponding to positive (negative) $\fNL$ and $b_1$.}
    \label{fig:GWDNeff_PBH}
\end{figure}

We can expand the compaction function \cref{eq:C1expl} in terms of short and long modes
\begin{align}
\label{eq:C1explaa}
\mathcal{C}(\vec x,r_m)
&
\simeq \mathcal{C}^s_1(r_m)\left(1+\frac{F_{\rm s}''}{F_{\rm s}'}\zeta_{{\rm g},l}(\vec x)\right)
\nonumber \\ &
-\frac{3}{8}\Big(\mathcal{C}^s_1(r_m)\Big)^2 \left(1+2\frac{F_{\rm s}''}{F_{\rm s}'}\zeta_{{\rm g},l}(\vec x)\right).
\end{align}
where we introduced $F_{\rm s} \equiv F(\zeta_{\rm s})$, and
\begin{equation}
\mathcal{C}^s_1(r_m)=-\frac{4}{3}r_m
F'_{\rm s}
\zeta_{{\rm g},s}^{\prime}(r_m),
\end{equation}
and primes denote derivatives evaluated at $\zeta_{{\rm g},l}=0$, and radial derivatives of the long mode are neglected as they are strongly suppressed.
Following Ref.~\cite{Crescimbeni:2025ywm} closely, one can neatly proceed by redefining a Gaussian component of the compaction function
$\mathcal{C}_1^s(r_m)
    \to 
    \left(
    1+\zeta_{{\rm g},l}(\vec x) F_{\rm s}''/F^\prime_{{\rm s}}\right)
    \mathcal{C}_1^s(r_m)$, 
along with the corresponding changes described in App.~\ref{app:PBH_formation} and the definitions of \cref{eq:Var1,eq:Var2}, to obtain
\begin{align}
\dpbh({\vec x})
    \simeq
    &\left(
    2\frac{\partial\ln P(\mathcal{C} > \mathcal{C}_{\rm th})}{\partial\ln \sigma_{\rm c}^2}
    \right . 
    \nonumber \\
    &\ \ \left .
    +
    \frac{\partial\ln P(\mathcal{C} > \mathcal{C}_{\rm th})}{\partial\ln \sigma_{\rm cr}^2}\right)
    \frac{F_{\rm s}''}{F_{\rm s}'}
    \zeta_{{\rm g},l} (\vec x).    
\end{align}
Following the standard notation in the literature, it is convenient to define the linear bias as a simple linear proportionality,  
\begin{equation}
    \delta_{\rm PBH}= b_1 \, \zeta_{{\rm g},l},
\end{equation}
which gives
\begin{multline}
    b_1 = \frac{1}{P({\cal C}>{\cal C}_{\rm th})} 
    \int_{\mathcal{D}} 
    \left(
    \frac{\mathcal{C}_g^2 \sigma^2_{\rm  r} - \mathcal{C}_g \zeta_{\rm g} \sigma^2_{\rm cr} }
    {\sigma^2_{\rm c} \sigma^2_{\rm  r} - \sigma^4_{\rm cr}}-1 
    \right )
    \\
  \times   \frac{F_{\rm s}''}{F_{\rm s}'} 
    \textrm{P}_{\rm g} (\mathcal{C}_g ,\zeta_{\rm g})
    \td\mathcal{C}_1 \td\zeta_{\rm g},
\end{multline}
where $\textrm{P}_{\rm g}$ and $\mathcal{D}$ (the integration domain where Hubble patches collapse to form PBHs) are defined in \cref{eq:PDFCompa} and we assumed a monochromatic mass distribution.
For simplicity, in the following we assume the non-linear relation of the usual form 
\begin{equation}
    F(\zeta_{\rm g}) = \zeta_{\rm g} + \frac{3}{5} f_{\rm NL} \zeta_{\rm g}^2.
\end{equation}
As a representative example, throughout the paper, we will assume a peaked power spectrum of enhanced curvature perturbations with a broken power-law shape with slopes $k^4$ in the IR and $k^{-3}$ in the UV, see more details in \cref{app:PBH_formation}.
We can finally show the linear bias as a function of $f_{\rm NL}$ for different values of $\beta$ in \cref{fig:GWDNeff_PBH}.
As expected, the linear bias is stronger for larger values of $f_{\rm NL}$ which indicates stronger couplings between long and short modes. Additionally, the linear bias grows for smaller $\beta$. This is because smaller $\beta$ corresponds to rarer collapse events, in the exponential tail of the PDF, which is then more sensitive to external effects from the coupling to the long modes. The difference between positive and negative $f_{\rm NL}$ comes from the non-linear relation curvature and the compaction function. We also checked that in the small $f_{\rm NL}$ limit, one finds $b_1 \propto f_{\rm NL}$, while the linear scaling is broken already at $\fNL \sim {\cal O}(1)$ due to non-linearities \cite{Crescimbeni:2025ywm}.

\subsection{PBH bias}

The dark matter density parameter $\Omega_{\rm DM}$, to first order in $\zeta$, can be expressed in terms of the background density parameter $\bar{\Omega}_{\rm DM}$ as
\begin{equation}
    \Omega_{\rm DM} = \left( 1 + f_{\rm PBH} \, b_1 \, \zeta + 3\zeta \right) \bar{\Omega}_{\rm DM}.
\end{equation}
Here, the $3\zeta$ term corresponds to the adiabatic mode to first order in $\zeta$.  
The term proportional to $f_{\rm PBH} b_1 \zeta$ represents an isocurvature contribution, leading to a deviation from the purely adiabatic case.

On CMB scales, observational constraints give \cite{Planck:2018jri}
\begin{equation}\label{iso:bound}
100 \beta_{\rm iso} <
\begin{cases}
0.095, & \text{fully correlated}, \\
0.107, & \text{fully anti-correlated},
\end{cases}
\end{equation}
where fully correlated modes correspond to $b > 0$, and fully anti-correlated modes correspond to $b < 0$. Notice we take the much stronger limits which apply to (anti)correlated perturbations as large isocurvature can only be induced by non-Gaussianities relating long scales to small perturbations, differently from what would happen in the case of Poisson-induced perturbations. 
The parameter $\beta_{\rm iso}$ can be expressed as
\begin{equation}
    \beta_{\rm iso} = \frac{P_{\rm iso}}{P_{\rm iso} + P_{\zeta}},
\end{equation}
where $P_{\rm iso}$ and $P_{\zeta}$ denote, respectively, the isocurvature and adiabatic curvature perturbation power spectra. These are related by $P_{\rm iso} = b_1^2 P_{\zeta}$.
If PBH, or the DM particle emitted by PBH evaporation, constitute a fraction $f_{\rm PBH}$ of the total dark matter content, the limits given in \cref{iso:bound} can be applied to place bounds on $b_1$ as  
\begin{equation}
\label{eq:bound b1 f}
-0.033 < b_1 f_{\rm PBH} < 0.031.
\end{equation}
Given the observational bound in Eq.~\eqref{eq:bound b1 f}, we see that values of $\fNL \gtrsim {\cal O}(10^{-3})$ rule out the entirety of DM being composed of a DS produced by evaporated PBHs, given the assumption of a scale-invariant non-Gaussianity.

\section{Results}
\label{sec:results}
\begin{figure}[t]\centering
\includegraphics[width=\columnwidth]{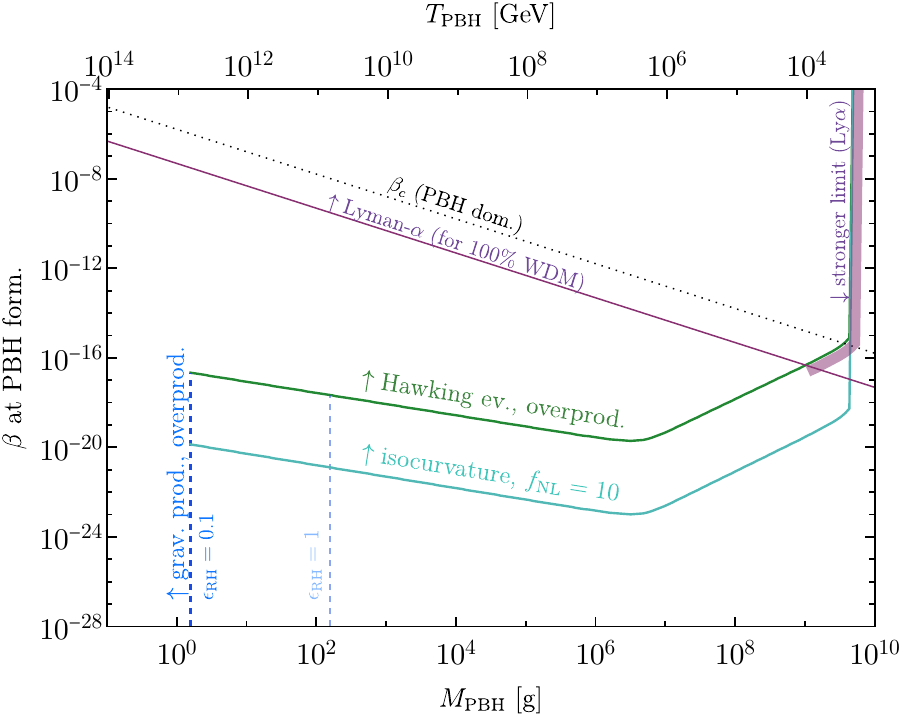}
\caption{Simplified version of \cref{fig:DM_PBH}, for the case $\mDM =10^7\GeV$, to highlight the main features of the bounds. In green we show the overproduction bound, in cyan the more stringent bound from isocurvature perturbations assuming $\fNL = 10$. Above the purple diagonal curve, all the DM being in the DS from evaporated PBHs is excluded by Lyman-$\alpha$ bounds on warm DM, while points on the left of the vertical dashed lines are excluded by gravitational overproduction, depending on the assumed $\epsilon_{\rm RH}$.  }
\label{fig:moneyplot_pedagogical}
\end{figure}

\begin{figure*}[h]\centering
\includegraphics[width=\linewidth]{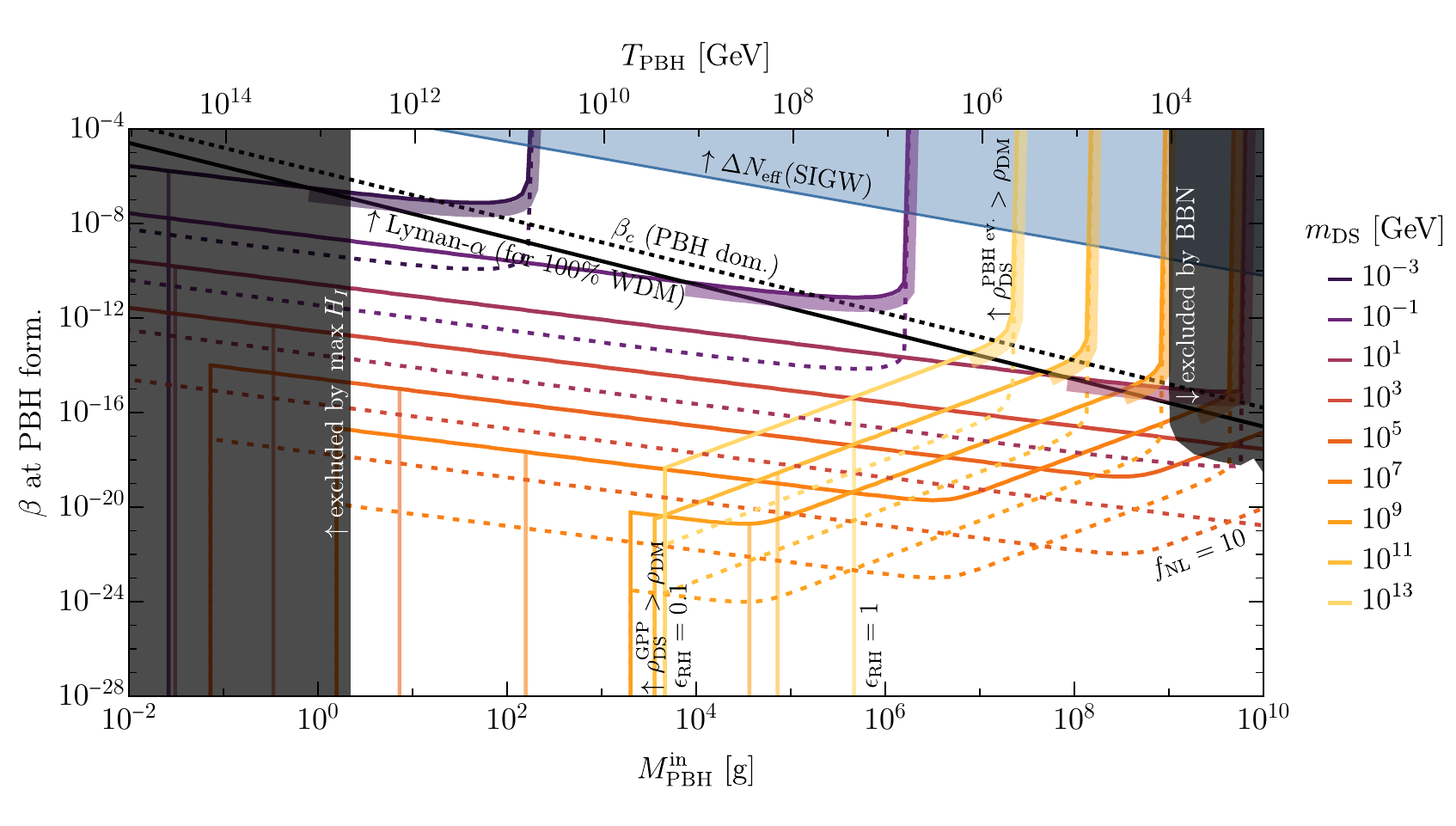}\vspace{-2em}
\caption{Values of $\beta$ , for given PBH mass $\MBH^{\rm in}$ yielding the observed DM abundance for various values of $\mDM$. Each contour corresponds to $\Omega_{\rm DM} h^2 = 0.11$; points above (below) overproduce (underproduce) dark matter.
Gray shaded regions are excluded by the maximum allowed $H$ (left), given the upper bound on the tensor-to-scalar-ratio $r$ (the PBH mass at formation being associated to the horizon mass), and by energy injection from evaporation at BBN (right). The triangular upper blue shaded region is excluded by SIGW emission overcoming the $\Delta N_{\rm eff}$ bound on the GW abundance.
}
\label{fig:DM_PBH}
\end{figure*}

\begin{figure*}[h]\centering
\includegraphics[width=\linewidth]{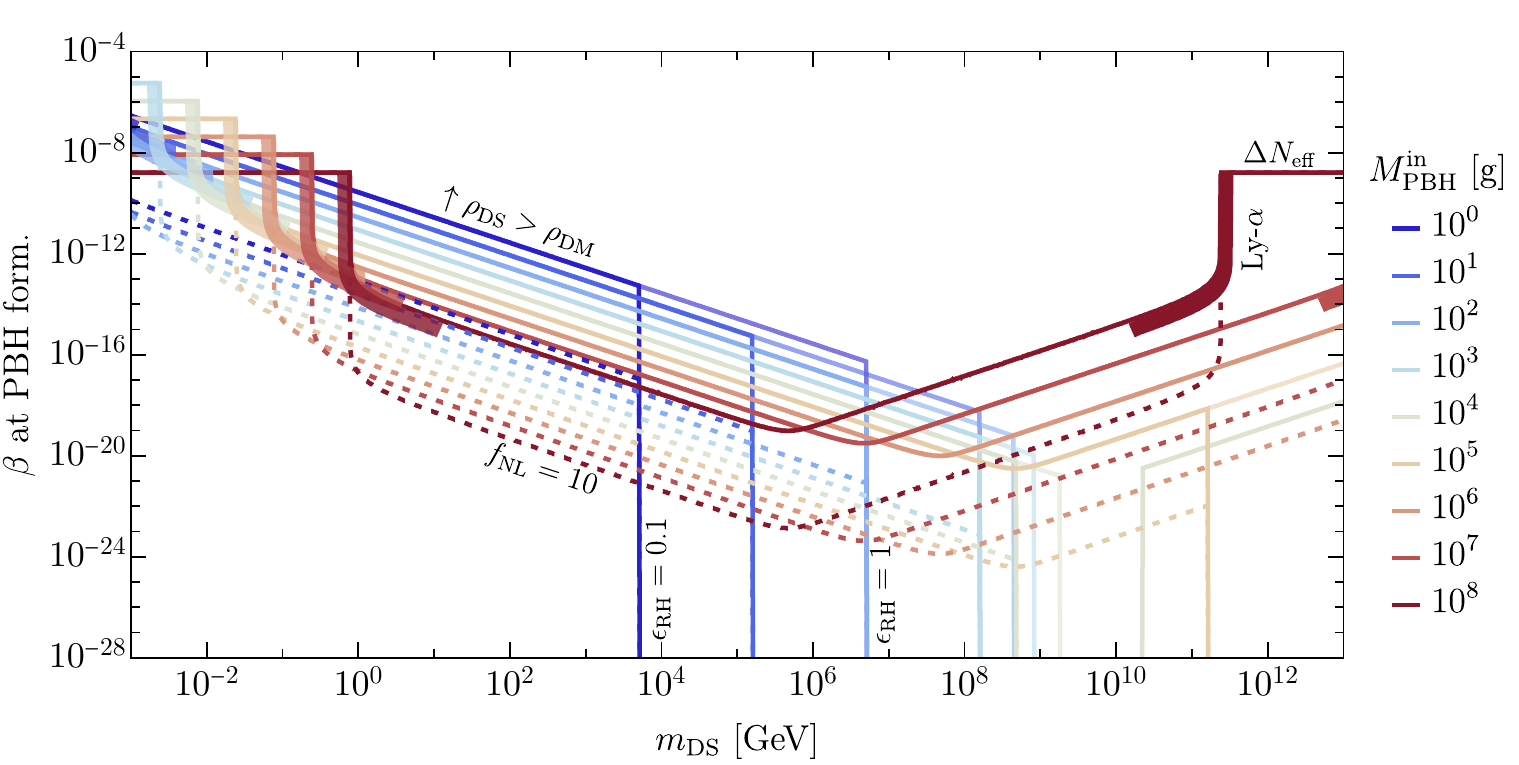}
\caption{ Same as \cref{fig:DM_PBH} but now showing the constraints as a function of $\mDM$ for different PBH masses.}
\label{fig:DM_PBH_2}
\end{figure*}

The main results of our paper are summarised in \cref{fig:DM_PBH,fig:DM_PBH_2}, which show two different projections of the constraints on the 3-dimensional parameter space $(\MBH^{\rm in}, \beta,\mDM)$, where $\mDM$ is the mass of a stable particle in the DS (whose abundance is not reprocessed later). For concreteness, we assume that such particles are fermions, but for everything we discuss (except inflationary gravitational production) the spin of the particle leads to $\mathcal O(1)$ numerical changes.
\cref{fig:DM_PBH} shows various constraints for fixed $\mDM$, and \cref{fig:DM_PBH_2} shows the envelope of these bounds for fixed $\MBH$. 
The main features of the contours in \cref{fig:DM_PBH} are reproduced in a simplified plot in \cref{fig:moneyplot_pedagogical} (for fixed $\mDM$), and are described in the following.

\noindent\textbf{$\bullet$ DM overproduction from Hawking evaporation}\quad

The correct DM abundance through Hawking evaporation is shown by the solid lines in \cref{fig:DM_PBH,fig:DM_PBH_2}. 
Points above each contour correspond to DM overproduction ($\Omega_{\rm DM} h^2 >$ 0.12), while those below lead to underproduction.
The main features of the contours in \cref{fig:DM_PBH,fig:moneyplot_pedagogical} can be understood as follows.

\paragraph{Euristic interpretation of the bounds}
It is instructive to reproduce the scaling of the bound with the dimensionful parameters $(\MP,\MBH^{\rm in},\mDM)$.
We begin by estimating $\mathcal N_i$, the total number of emitted particles per PBH for the DS species $i$. 
We integrate the rate \cref{eq:BHrate} (in this paragraph, we omit all dimensionless prefactors),
\begin{equation}
\mathcal N_i \sim \int_{t_f}^{t_\text{ev}} \mathrm dt \int_0^\infty \mathrm dp \frac{\sigma}{e^{E/\TBH(t)}\pm 1} \frac{p^3}{E} \,.
\end{equation}
The absorption cross section $\sigma$, up to greybody factors, is of the order of the geometric cross section $r_s^2 \sim \MBH^2(t)/\MP^4$. 
The phase space distribution $1/(e^{E/\TBH(t)}\pm 1)$ yields a prefactor $\sim\mathcal O(1)$ if the $i$ particle is lighter than the initial Hawking temperature, $\mDM < \TBH(t_f)$, while it implies a shutoff of the production as long as $\max(\mDM,p) > \TBH(t)$ in the opposite case. The momentum integral can be estimated as
\begin{multline}
\int_0^{\TBH(t)} 
\mathrm dp\, p^2 \frac{\MBH^2(t) }{\MP^4}
e^{-\frac{\mDM}{\TBH(t)}}
\sim
\frac{\MP^2 e^{-\frac{\mDM}{\TBH(t)}}
}{\MBH(t)}.
\end{multline}
We now change the integration variable with \cref{eq:dMBHdt}, $\mathrm dt \to -\mathrm d\MBH(\MBH^2/\MP^4)$. 
We can see that particle production from Hawking evaporation is an IR-dominated process, as the production rate peaks at final stages ($\propto 1/\MBH(t)\sim \TBH(t)$) but the evaporating PBH spends a longer time around the initial mass ($\mathrm dt \sim \MBH^2 \mathrm d\MBH$), with the latter effect prevailing.
The time (or mass) integral is then dominated by the largest BH mass $M_{\max}=\min(\MBH^\text{in},\MP^2/\mDM)$ when the production is not Boltzmann suppressed (which occurs if $\mDM>\TBH(t) \sim \MP^2/\MBH(t)$):
\begin{equation}
\mathcal N_i \sim \int_0^{M_{\max}}\hspace{-1.8em} \mathrm d \MBH \frac{\MBH}{\MP^2} \sim \min\left( \frac{(\MBH^\text{in})^2}{\MP^2},\frac{\MP^2}{\mDM^2}\right) 
\end{equation}
This is a key ingredient to compute the value of $\beta$ giving the correct DM abundance at equality: 
\begin{equation}\label{beta_step}
\beta \sim \frac{\MBH^{\rm in} n_\PBH}{T_f^4} 
    \sim \frac{\MBH^{\rm in}\, \rho_\text{DM,eq}}{\mathcal N_i \mDM T_f T_\text{eq}^3}
\end{equation}
where $T_f$ is the temperature of the thermal bath at PBH formation, and $\rho_\text{DM,eq}/T_\text{eq}^3$ is a fixed quantity at equality. In this estimate for $\beta$, we focus on the DS particles produced right after the PBH formation (or as soon as they are not Boltzmann suppressed), so everything is evaluate at PBH formation via \cref{eq:Min}: $H_\times\sim \MP^2/\MBH^{\rm in}$, $T_f \sim \sqrt{H_\times \MP}$.
We can then write the final scaling of the bound on $\beta$ \cref{beta_step} with $(\MBH^{\rm in},\mDM)$  as 
\begin{equation}\label{eq:beta euristic}
\beta \sim \left\lbrace
\begin{aligned}
&(\MBH^{\rm in})^{-1/2} \mDM^{-1} && \text{if } \mDM < \TBH^\text{in},\\
&(\MBH^{\rm in})^{3/2} \mDM && \text{if } \mDM > \TBH^\text{in}.
\end{aligned}
\right.
\end{equation}
We can now cross-check this prediction with the numerics.


\paragraph{Small PBH masses}
In the limit $\MBH^{\rm in} \to 0$, the Hawking temperature satisfies $\TBH \gg \mDM$, ensuring efficient DM production throughout the entire evaporation process. 
This regime corresponds to the descending curves on the left side of \cref{fig:DM_PBH}.

\paragraph{Intermediate masses} For larger PBH masses $\mDM>\TBH^\text{in}$, DM production is Boltzmann-suppressed. 
This results in a rising trend in the contours for $\mDM \gtrsim 10^5~{\rm GeV}$, where the contours develop a noticeable "knee".

\paragraph{PBH domination regime}: If the initial PBH abundance satisfies $\beta > \beta_c \equiv T_{\rm ev}/T_{\rm in}$, PBHs come to dominate the energy density of the Universe before they evaporate. 
In this case, the final DM abundance no longer depends on $\beta$, but only on the PBH mass. This is because both the DM and SM radiation are produced via Hawking evaporation, and their relative contributions to the energy budget are set solely by the properties of the emission process. This behavior is reflected in the plot by vertical contour lines beyond the threshold $\beta = \beta_c$ (shown with a black dotted line). For sufficiently large PBH masses in this regime, the correct relic abundance cannot be achieved even for arbitrarily high values of $\beta$, causing the contours to turn sharply upward.

The numerical fit for the final bound on $\beta$ confirms the analytical estimate of \cref{eq:beta euristic}:
\begin{widetext}
\begin{equation} \label{eq:beta fit}
\begin{aligned}
\beta < \max\bigg[ &
    2.8\cdot 10^{-10}\parfrac{\MBH^{\rm in}}{\g}^{-1/2} \parfrac{\mDM}{\GeV}^{-1}
    , 
    & \text{(small $\MBH^{\rm in}$)}\\ & 
    1.7\cdot 10^{-37}\parfrac{\MBH^{\rm in}}{\g}^{3/2} \parfrac{\mDM}{\GeV}
    ,
    & \text{(intermediate $\MBH^{\rm in}$)} \\& 
    \Theta\left( \frac{\MBH^{\rm in}}{\g} - 3.7\cdot 10^{12}\frac{(\mDM/\GeV)^{-0.4}}{1+3.5\cdot 10^4(\mDM/\GeV)^{-2.3}} 
    \right) 
    \bigg]  
    & \text{(PBH domination)} 
\end{aligned}
\end{equation}
\end{widetext}
where $\Theta(\cdot)$ is the Heavyside theta function.
This formula captures very effectively the envelope of the decreasing curve on the left-hand side of \cref{fig:DM_PBH} (with a slope $(\MBH^{\rm in})^{-1/2}$), the increasing curve on the right-hand side (for $\mDM\gtrsim 10^5 \GeV$, with a slope $(\MBH^{\rm in})^{3/2}$), and the endpoint when the curve trespasses the boundary for PBH domination.

\noindent\textbf{$\bullet$ Isocurvature}\quad
In this paper we compute also the bound from the isocurvature perturbations displayed by PBHs in presence of primordial NG coupling long and short modes, as discussed in \cref{sec:Iso PBH NG}. This is shown in \cref{fig:DM_PBH,fig:DM_PBH_2} by the dashed coloured lines, corresponding to the choice $\fNL=10$.
In the presence of NG, the isocurvature perturbation forces the DS particles produced by the evaporation to be a subcomponent of DM, according to the \cref{eq:bound b1 f}, which makes the bounds in $\beta$ stronger.
This equation, combined with \cref{eq:beta fit} and with the result of  \cref{fig:GWDNeff_PBH}, allows to draw the bound on PBH evaporation from isocurvature for a given $\fNL$ and $\mDM$. 
As a reference, $\fNL=1$ would lead to an intermediate constraint between $\fNL=0$ and $\fNL=10$.

\noindent\textbf{$\bullet$ DM overproduction from gravitational production}\quad
As discussed in \cref{sec:DM from GPP}, for given inflationary Hubble rate $\HI$ and maximum temperature $\Tmax$ after reheating, there is a bound from the gravitational  particle production (GPP) of any stable particle. We show it in the left panel of \cref{fig:DM GPP and UV FI}. In the right panel, we relate $\HI$ to $\Tmax$ via the parameter $\epsRH = \Tmax/\TRH$, and we connect $\Tmax$ to the minimum $\MBH$ that can be produced (which corresponds to perturbations sourced at the end of inflation).

As a result, for given $\mDM$ and $\epsRH$, there is a minimum PBH mass that is allowed: a lower value would imply a too large $\HI$ or $\Tmax$ which would overproduce the DS particle. This lower limit is shown by vertical lines for $\epsRH=0.1$ (and $\epsRH=1$ with a lighter colour) for each $\mDM$.

\noindent\textbf{$\bullet$ Warm DM}\quad
We reviewed in \cref{sec:WDM} the constraint from Lyman-$\alpha$ observations on the lack of power suppression on short scales, which would be caused if DM were warm. This is captured by the bound shown by a solid black line, which assumes no other DM component besides the DS particle produced by Hawking evaporation. 
We mark this bound in \cref{fig:DM_PBH} as a thick shading extending the coloured lines for a given $\mDM$.
In reality, a more accurate calculation of this bound allowing the warm component to be a fraction of the total DM would allow to reinforce the exclusion contours in the vicinity of the coloured shadings. As this task goes beyond the scope of the present paper, we only mark visually with a thick shading where the bounds would be strengthened.

\noindent\textbf{$\bullet$ Scalar-induced GWs}\quad
The upper bound on $\beta$ from the overproduction of scalar-induced GWs (and the consequent increase of $\Neff$, constrained by BBN) is shown by a blue line in the upper right corner of \cref{fig:DM_PBH}. This bound translates into the upper horizontal lines of \cref{fig:DM_PBH_2}.

\cref{fig:DM_PBH_2} shows the projection of the constraints to the parameter space $(\mDM, \beta)$ for different values of $\MBH$. This is the natural parameter space to consider for a given DS with stable particles of mass $\mDM$: the coloured lines then show the maximum initial abundance of PBHs with a given mass.
The vertical (cutting-off) lines, related to inflationary gravitational production, can be understood as a lower limit on the PBH mass that can be generated for the largest $\HI$ avoiding overproduction of the DS particle.

\section{Conclusions}
\label{sec:conclusions}

We have revisited the production of stable dark sector particles from the complete evaporation of light PBHs, focusing in particular on the possible cosmological implications of isocurvature perturbations. We have shown that, although PBHs are seeded by Poisson fluctuations on microscopic scales, primordial non-Gaussianity is required to couple long- and short-wavelength modes, thereby generating PBH number density fluctuations on cosmological scales. As a consequence, DS particles produced through Hawking evaporation inherit isocurvature perturbations whose amplitude is controlled by the PBH bias and by the fraction of DM originating from PBHs.

We have quantified the impact of current CMB limits on correlated and anti-correlated DM isocurvature, translating them into bounds on the PBH abundance and on the parameter space $(M_{\rm PBH}^{\rm in},\beta,m_{\rm DS})$. In addition, we have reassessed the allowed region of this scenario by consistently combining constraints from DM overproduction via Hawking evaporation, gravitational particle production during and after inflation, warm DM limits from Lyman-$\alpha$ observations, and the bound on scalar-induced gravitational waves contributing to $\Delta N_{\rm eff}$. Our results provide a comprehensive mapping of the viable parameter space and highlight the complementarity among these probes.

A key prediction of this framework is that any isocurvature signal generated through PBH evaporation is expected to be fully correlated or anti-correlated with the adiabatic curvature perturbations, reflecting its origin from primordial non-Gaussianity. This property could allow future observations to distinguish this scenario from alternative DM production mechanisms that may generate uncorrelated or Poisson-like isocurvature perturbations (e.g. axion DM \cite{Lyth:1991ub, Fox:2004kb, Hertzberg:2008wr, Visinelli:2009zm, Chen:2023txq,Graham:2025iwx}). Furthermore, in parts of the viable parameter space the DM component produced via PBH evaporation can possess a mildly warm spectrum, potentially leaving observable imprints in future small-scale structure measurements.

Several directions deserve further investigation. A refined treatment of warm DM bounds including mixed cold--warm DM compositions would sharpen the constraints derived here. A more detailed modelling of non-Gaussianity beyond the local ansatz could improve the robustness of clustering and isocurvature predictions. Finally, connecting the required PBH formation spectra to explicit inflationary models would clarify the microphysical origin of the scenario and its associated GW signatures. Together, these developments could strengthen the testability of dark sector production from PBH evaporation in upcoming cosmological and astrophysical observations.

\begin{acknowledgments}
\vspace*{-1.em}%
\begin{small}\noindent
We thank Marco Chianese, Isabella Masina and Jessica Turner for useful comments on the draft.\\
G.F.~acknowledges support by the
Italian MUR Departments of Excellence grant 2023–2027
“Quantum Frontier” and from Istituto Nazionale di Fisica Nucleare (INFN) through the Theoretical Astroparticle Physics (TAsP) project.\\
D.R.~is supported by the ``Rita Levi-Montalcini'' programme for young researchers of MUR, and was supported at U.~of Zurich by the UZH Postdoc Grant 2023 Nr.\,FK-23-130.
\end{small}
\end{acknowledgments}

\appendix 
\onecolumngrid
\section{PBH formation basics}
\label{app:PBH_formation}

In this section, we summarize the formalism used to compute the PBH abundance and define some key quantities used in the main text. We follow state-of-the-art modeling of the collapse probability, using the notation described in e.g.~\cite{Ferrante:2022mui}.\footnote{For a public code when assuming the absence of non-Gaussianities, see Refs.~\cite{Cecchini:2025oks} and \cite{fastPTAmodule}.}
We report the computation adopting threshold statistics~\cite{Young2025}, but acknowledge that systematic uncertainties remain, as peak-theory computations~\cite{Bardeen:1985tr} indicate that lower perturbation amplitudes are needed~\cite{Young:2014ana,DeLuca:2019qsy,Iovino:2024tyg,Cecchini:2025oks}. This, however, does not affect our conclusions as far as clustering properties is concerned, but only concerns the connection of the PBH scenario with specific formation models. For the role of non-Gaussianities in the computation of $f_\PBH$, see~\cite{Yoo:2018kvb,Byrnes:2012yx,Atal:2018neu, DeLuca:2022rfz, Franciolini:2018vbk,Ferrante:2022mui,DeLuca:2019qsy,Young:2019yug,Biagetti:2021eep,Gow:2022jfb}.

\paragraph{Compaction function}
The PBH formation criterion and clustering properties rely on the definition of a compaction function~\cite{Shibata:1999zs}, computed as twice the mass excess over the background and divided by the areal radius $R(r,t) = a(t)\, e^\zeta\, r$. This means
\begin{align}\label{eq:DefinitionCompaction}
\mathcal{C}(r,t) = \frac{2\left[M(r,t) - M_b(r,t)\right]}{R(r,t)} =
\frac{2}{R(r,t)}\int_{V_{R}} d^{3}\vec{x}\,
\rho_b(t)\delta(\vec{x},t).
\end{align}
At early times, which set the initial conditions for the evolution of overdensities at super-Hubble scales, and assuming spherical symmetry, the density contrast reads~\cite{Harada:2015yda}
\begin{equation}\label{eq:SphericalDelta}
\delta(r,t) = 
-\frac{4}{9}
\left(\frac{1}{aH}\right)^2 
e^{-2\zeta(r)}\left[
\zeta^{\prime\prime}(r) + \frac{2}{r}\zeta^{\prime}(r) + \frac{1}{2}\zeta^{\prime 2}(r)
\right],
\end{equation}
where $' \equiv d/dr$. The numerical prefactor assumes radiation domination. In case PBHs are produced in different scenarios, such as early matter phases, one needs to adjust the eqaution accordingly (e.g.~\cite{Harada:2016mhb}). Plugging this into \cref{eq:DefinitionCompaction}, one finds~\cite{Harada:2015yda}
\begin{equation}\label{eq:CompactionFull}
\mathcal{C}(r) = 
-\frac{4}{3}\,r\,\zeta^{\prime}(r)\left[
1 + \frac{r}{2}\zeta^{\prime}(r)
\right] = 
\mathcal{C}_1(r) - \frac{3}{8}\mathcal{C}^2_1(r),
\qquad \text{where}
\qquad
\mathcal{C}_1(r) = -\frac{4}{3} r\zeta^{\prime}(r).
\end{equation}
The scale $r_m$ where $\mathcal{C}$ peaks satisfies
$\mathcal{C}^{\prime}(r_m) = 0 $
which solves the equation
$\zeta^{\prime}(r_m) + r_m\zeta^{\prime\prime}(r_m) = 0$. For the typical shape of overdensities, the compaction function peaks at $r_m \equiv \kappa /k$ with $\kappa = {\cal O}(3)$ depending on the peak profile \cite{Musco:2020jjb}, and where $k$ is the comoving wavenumber.
PBH formation requires $\mathcal{C}_{{\rm max}} = \mathcal{C}(r_m) > \mathcal{C}_{\rm th}$, where the latter is determined by dedicated numerical simulations \cite{Musco2025}. At horizon crossing $r_m = (aH)^{-1}$, $\mathcal{C}_{{\rm max}}$ corresponds to the smoothed nonlinear density contrast.

When introducing non-Gaussianities of the form $\zeta = F(\zeta_{\rm g})$, the linear compaction becomes
\begin{align}\label{eq:C1expl}
\mathcal{C}_1(r) = -\frac{4}{3}\,r\,\zeta_{\rm g}^{\prime}(r)\,
\frac{dF}{d\zeta_{\rm g}} = 
\mathcal{C}_{\rm g}(r)\,
\frac{dF}{d\zeta_{\rm g}},
\qquad \text{where} \qquad
\mathcal{C}_{\rm g}(r) =
-\frac{4}{3}r\zeta_{\rm g}^{\prime}(r),
\end{align}
leading to
$\mathcal{C}(r) = 
\mathcal{C}_{\rm g}(r)\,
F' - {3}
 \mathcal{C}^2_{\rm g}(r)
 \left(F'\right)^2$/8 ,
where $F'$ denotes $dF/d\zeta_{\rm g}$. Notice that now both $\zeta_{\rm g}$ and $\mathcal{C}_{\rm g}$ are Gaussian, and can be 
expressed as filtered fields
\begin{align}
\mathcal{C}_{\rm g}(r) &= -\frac{4}{9}r^2\int \td^3y\,\nabla^2\zeta_{\rm g}(\vec y)\,W(\vec x-\vec y,r),
\qquad 
\zeta_{\rm g}(r) = \int \td^3y \,\zeta_{\rm g}(\vec y)\,W_{\rm s}(\vec x-\vec y,r),
\end{align}
where, now expressed in momentum space, $\tilde W_{\rm s}(k,r) = \sin(kr)/kr$ and $\tilde W(k,r) = 3[\sin(kr) - kr\cos(kr)]/(kr)^3$.

\paragraph{Collapse probability}
The joint Gaussian PDF takes the simple form
\begin{align}\label{eq:PDFCompa}
 \textrm{P}_{\rm g}(\mathcal{C}_{\rm g},\zeta_{\rm g}) 
 = \frac{1}{(2\pi)\sqrt{\det\Sigma_1}}
 \exp\left(
 -\frac{1}{2}\vec{Y}_1^{\rm T}\Sigma_1^{-1}\vec{Y}_1
 \right),
\end{align}
with $\vec{Y}_1 = (\mathcal{C}_{\rm g}, \zeta_{\rm g})^{\rm T}$ and covariance $\Sigma_1$ takes diagonal elements
\begin{align}
\sigma_{\rm c}^2 = \frac{16}{81}\int_0^{\infty}\frac{\td k}{k}(kr_m)^4 \tilde W^2(k,r_m) T^2(k, r_m) P_\zeta(k),
\qquad
\sigma_{\rm  r}^2 = \int_0^{\infty}\frac{\td k}{k}\tilde W_{\rm s}^2(k,r_m) T^2(k, r_m)  P_\zeta(k),\label{eq:Var1}
\end{align}
and off-diagonal elements
\begin{align}
    \sigma_{\rm cr}^2 &= \frac{4}{9}\int_0^{\infty}\frac{\td k}{k}(kr_m)^2 \tilde W(k,r_m) \tilde W_{\rm s}(k,r_m) T^2(k, r_m) P_\zeta(k).\label{eq:Var2}
\end{align}
In the previous equations, we introduced the linear transfer function $T(k,r_m)$ describing the evolution of curvature perturbations in linear cosmological perturbation theory. Additionally, $P_\zeta$ is the dimensionless curvature power spectrum.

The probability for collapse, which is $\mathcal{C} > \mathcal{C}_{\rm th}$,  is now computed using threshold statistics as 
\begin{align}
     P(\mathcal{C} > \mathcal{C}_{\rm th}) =
    \int_{\mathcal{D}} \textrm{P}_{\rm g}
    (\mathcal{C}_{\rm g} ,\zeta_{\rm g})\,
    \td\mathcal{C}_{\rm g} \td\zeta_{\rm g},
\qquad \text{where} \qquad 
\mathcal{D} = \left\{
\mathcal{C}_{\rm g},\,\zeta_{\rm g} \in \mathbb{R}:
\mathcal{C} > \mathcal{C}_{\rm th},\,
\mathcal{C}_1 < 4/3
\right\}.
\end{align}

\paragraph{PBH mass and mass fraction}
The PBH mass relates to the horizon mass via the scaling law~\cite{Choptuik:1992jv,Evans:1994pj}
\begin{equation}
    M_{\rm PBH}(\mathcal{C}) = \mathcal{K} M_{\rm H} (\mathcal{C} - \mathcal{C}_{\rm th})^{\gamma}, \quad \gamma = 0.38,
\qquad 
\text{
and~\cite{Musco:2020jjb}
}
\qquad 
    M_{{\rm H}} \simeq 17 M_\odot\left(\frac{g_{\star}}{10.75}\right)^{-1 / 6}\left(\frac{k / \kappa}{10^6 \mathrm{Mpc}^{-1}}\right)^{-2}.
\end{equation}
Finally, the mass fraction $\beta$ is computed as
\begin{align}\label{eq:beta}
    \beta = 
    \int_{\mathcal{D}}
    \frac{M_{\rm PBH}({\cal C})}{M_{H}}
    {\rm P}_{g}(\mathcal{C}_{\rm g},\zeta_{\rm g})
    \td\mathcal{C}_{\rm g}\td\zeta_{\rm g}.
\end{align}
We use the threshold $\mathcal{C}_{\rm th} = 0.56$ from Ref.~\cite{Musco:2020jjb}, evaluated at $r_m$.

\twocolumngrid
\bibliography{main}

\end{document}